\def\HB{\overline\HH}
\def\RB{\overline R}
\def\cd{\mathop{\rm cd}}
\def\tr{\mathop{\rm tr}}
\def\dist{{\rm dist}}
\def\cd{{\rm cd}}
\def\twovec#1#2{\left[\matrix{#1\cr #2\cr}\right]}
%auto-ignore

\hd{Analysis of $\mu_2$} 
To analyze the function $\mu_{3,p}$ we will write it in terms of
$\mu_2$, defined by
$$\eqalign{
\mu_2(z_1,z_2,q_1,q_2,\lambda) 
&= {{2\cd(\phi(z_1,z_2,q_1,q_2,\lambda))}\over{\cd(z_1)+\cd(z_2)}}\cr
%&= {{2 \chi(z_1)\chi(z_2)}\over{\chi(\phi(z_1,z_2,q_1,q_2,\lambda))(\chi(z_1)+\chi(z_2))}}\cr
}$$
initially as a function from $\HH^2\backslash\{(z_\lambda,z_\lambda)\}\times \RR^2\times R\rarr \RR$. 
In this section $R=R(E,\epsilon)$ for some $0<E<2\sqrt{2}$ and $\epsilon>0$.
(Note that here and throughout this paper we are using $\HH^n$ to denote a product
of hyperbolic planes, and not $n$-dimensional hyperbolic space.)

\beginpropositionlabel{mu2le1}
For all $z_1,z_2\in \HH^2\backslash\{(z_\lambda,z_\lambda)\}$ and 
$\lambda\in R$,
$$
\mu_2(z_1,z_2,0,0,\lambda) < 1\,.
$$
\endproposition

\beginproof
For $z,w\in\HH$ set 
$$
{\rm c}(w,z) = 2(\cosh(\dist_\HH(w,z))-1) = {{|w-z|^2}\over{\Im(w)\Im(z)}}\,.
$$
Note that $z\mapsto {\rm c}(w,z)$ is strictly convex. This can be seen for
example by noting that its Hessian has strictly positive eigenvalues.
Also, ${\rm c}(w,z)$ is invariant under hyperbolic isometries. Thus
$$
{\rm c}(2w,z_1+z_2) = {\rm c}(w,{{z_1+z_2}\over{2}}) \le {{1}\over{2}} {\rm c}(w,z_1) + 
{{1}\over{2}} {\rm c}(w,z_2) \,.
$$
Substituting $-(z_1-\lambda)^{-1}$ for $z_1$ and $-(z_2-\lambda)^{-1}$ for $z_2$
yields
$$\eqalign{
{\rm c}(2w,\phi(z_1,z_2,0,0,\lambda)) 
&\le {{1}\over{2}} {\rm c}(w, -(z_1-\lambda)^{-1}) + {{1}\over{2}} 
{\rm c}(w, -(z_2-\lambda)^{-1})\cr
&= {{1}\over{2}} {\rm c}(2w, -2(z_1-\lambda)^{-1}) + {{1}\over{2}} 
{\rm c}(2w, -2(z_2-\lambda)^{-1})\,.\cr
}$$
Now choose $2w=z_\lambda$. Since $z_\lambda$ is the fixed point of $z\mapsto -2(z-\lambda)^{-1}$
we obtain
$$
\cd(\phi(z_1,z_2,0,0,\lambda)) \le {{1}\over{2}} \cd(z_1) + {{1}\over{2}} \cd(z_2)\,.
$$
If equality holds then strict convexity in the first estimate above implies
$z_1=z_2$. Then, since $\Im(\lambda)>0$, $z\mapsto \phi(z,z,0,0,\lambda)$ is a strict
contraction with fixed point $z_\lambda$ (see [FHS]). This implies that the common
value of $z_1$ and $z_2$ must be $z_\lambda$. 
\endproof

We need to understand the behaviour of $\mu_2(z_1,z_2,q_1,q_2,\lambda)$
as $z_1$ and $z_2$ approach infinity, and $\lambda$ approaches the real axis.
We know from \thmrf{mu2le1} that the value of $\mu_2$ is at most one, and wish
to determine at what points it is equals one.
Thus it is natural to introduce the compactification 
$\HB\,^2\times \RR^2\times \RB$. Here $\RB$ denotes
the closure, and $\HB$ is the compactification of $\HH$ obtained
by adjoining the boundary at infinity. (The word compactification is not quite accurate
here because of the factors of $\RR$, but we will use the term nevertheless.) 

The boundary at infinity is defined as follows.
Cover the upper half plane model of the hyperbolic plane  
$\HH$ with two coordinate patches, one where
$|z|$ is bounded below  and one where $|z|$ is bounded above. On the patch where $|z|>C$ 
we use the co-ordinate function $w=-1/z$. Each chart looks like a semi-circle
in the complex plane of the form
$\{z\in \CC : \Im(z) > 0, |z| < C\}$. The boundary at infinity
consists of the sets $\{\Im(z)=0\}$ and $\{\Im(w)=0\}$ in the respective charts.
The compactification $\HB$ is the upper half plane with the boundary at infinity adjoined.
We will use $i\infty$ to denote the point where $w=0$.

We now think of $\mu_2$ as being defined in the interior of the
compactification $\HB\,^2\times \RR^2\times \RB$ and ask how
it behaves near the boundary. 
It turns out that in the co-ordinates introduced above, $\mu_2$ is a rational function.
At most points on the boundary the denominator does not vanish in the limit, 
and $\mu_2$
has a continuous extension. There are, however, points where both numerator
and denominator vanish, and at these singular points the limiting value of $\mu_2$
depends on the direction of approach. By blowing
up the singular points, it would be possible to define a compactification of 
$\HH^2\times \RR^2\times R$ to which $\mu_2$ extends continuously. 
However, this is more than we need for our proof. We will do a 
partial resolution of the singularities of $\mu_2$, consisting of two
blow-ups of the simplest kind, and then extend $\mu_2$ to an upper
semi-continuous function on the resulting compactification.

The reciprocal of the function $\cd(z)$,
$$\chi(z) = 1/\cd(z) = {{\Im(z)}\over{|z-z_\lambda|^2}}$$ 
is a boundary defining function for $\HH$. This means that in
each of the two charts above, $\chi$ is positive near infinity and 
vanishes exactly to first order on the boundary at infinity.

We will now describe our compactification of $\HH^2\times \RR^2\times R$.
Start with $\HB\,^2\times \RR^2\times \RB$.
The first blowup consists of writing $\chi(z_1), \chi(z_2)$ in polar
co-ordinates. Thus we introduce new variables $r_1$, $\omega_1$ and
$\omega_2$ and impose the equations
\be{bu1.1}
\eqalign{
\chi(z_1) &= r_1\omega_1\,,\cr
\chi(z_2) &= r_1\omega_2\,,\cr
}\ee
and
\be{bu1.2}
\omega_1^2 + \omega_2^2 = 1\,.
\ee
The blown up space is the variety in 
$\HB\,^2\times \RR^2\times \RB\times\RR^3$
containing all points 
$(z_1,z_2,q_1,q_2,\lambda,r_1,\omega_1,\omega_2)$ 
that satisfy \rf{bu1.1} and \rf{bu1.2}.

In the region where $|z_1|$ and $|z_2|$ are bounded, we
could use $\chi(z_1)$, $\chi(z_2)$,
$\Re(z_1)$, $\Re(z_2)$, $q_1$, $q_2$, $\lambda$ as local co-ordinates for
the original space $\HB\,^2\times \RR^2\times \RB$. The image
of such a co-ordinate chart near the boundary
would be $[0,\epsilon)^2\times I^2 \times \RR^2 \times \RB$
for some interval $I$. Local co-ordinates for the blown up space 
would be $r_1$, $\theta$, $\Re(z_1)$, $\Re(z_2)$, $q_1$, $q_2$, $\lambda$
where $\omega_1=\cos(\theta)$ and $\omega_2=\sin(\theta)$.
The image of such a chart in the blown up space would be
$[0,\epsilon)\times [0,\pi/2] \times I^2 \times \RR^2 \times \RB$.
Similarly, we could write local co-ordinates in the other regions.
The singular locus for the first blowup is the corner 
$\PA_\infty(\HB)\times\PA_\infty(\HB)\times \RR^2\times \RB$
in $\HB\,^2\times \RR^2\times \RB$, defined
by $\chi(z_1) = \chi(z_2) =0$. Corresponding to each point
in the singular locus is a quarter circle of points in the blown
up space, parametrized by $\omega_1,\omega_2$. Away from the
singular locus the original space and the blown up space are 
essentially the same, since we can solve for $r_1,\omega_1,\omega_2$ 
in terms of the original variables.

For the second blowup we introduce an additional real variable $r_2$ and two
additional complex
variables $\eta_1$ and $\eta_2$. We impose
\be{bu2.1}\eqalign{
z_1+\Re(\lambda)-q_1 &= r_2\,\eta_1\,,\cr
z_2+\Re(\lambda)-q_2 &= r_2\,\eta_2\,,\cr
}\ee
with
\be{bu2.2}
|\eta_1|^2 + |\eta_2|^2 = 1\,,
\ee
and 
$$
r_2\ge 0\,.
$$ 
The variables of the first and second blowups are not independent when $r_1,r_2\ne 0$. In fact, since 
$\chi(z_1)=r_2\Im(\eta_1)/|r_2\eta_1-\Re(\lambda)+q_1-z_\lambda|^2 = r_1\omega_1$ we find that
$r_1 r_2 \Im(\eta_1)\omega_2|r_2\eta_2-\Re(\lambda)+q_2-z_\lambda|^2$ and 
$r_1 r_2 \Im(\eta_2)\omega_1|r_2\eta_1-\Re(\lambda)+q_1-z_\lambda|^2$ are equal
so that, when $r_1,r_2\ne 0$,
\be{bu2.3}
\Im(\eta_1)\omega_2|r_2\eta_2-\Re(\lambda)+q_2-z_\lambda|^2= \Im(\eta_2)\omega_1|r_2\eta_1-\Re(\lambda)+q_1-z_\lambda|^2\,.
\ee
We will require that this equation be satisfied everywhere. Otherwise, there would
be points in the blown up space (where $r_2=0$ and \rf{bu2.3} is not satisfied) 
that are not in the closure of the interior of the original space.

As before, the twice blown up space is essentially the same as the once blown up
space away from the singular locus $z_1=-\Re(\lambda)+q_1$, $z_2=-\Re(\lambda)+q_2$.
Local co-ordinates for the twice blown up space near the singular locus are given
by $r_2$, $\omega_1$, $\omega_2$, $\Re(\eta_1)$, $\Re(\eta_2)$, $q_1$, $ q_2$, $\lambda$.

Define $K$ to be the space obtained from $\HB\,^2\times \RR^2\times \RB$
by the two blowups described above. The topology is the one given by the local
description as a closed subset of Euclidean space.
The boundary at infinity is defined to be
$$
\PA_\infty K = \{\chi(z_1)=0\} \cup \{\chi(z_2)=0\} = \{r_1=0\} \cup \{\omega_1=0\} \cup \{\omega_2=0\}\,.
$$
The set $K\backslash \PA_\infty K$ can be identified with $\HH^2\times \RR^2\times \RB$.

Extend $\mu_2$ to an upper semi-continuous function on $K$ by defining, for points $k\in\PA_\infty K$,
$$
\mu_2(k) = \limsup_{{k_n\rarr k}\atop{k_n\in K\backslash \PA_\infty K}}\mu_2(k_n)\,.
$$
Here $k_n\rarr k$ means convergence in $K$. More explicitly,
$k_n$ is a point $(z_{1,n},z_{2,n},q_{1,n},q_{2,n},\lambda_n) \in \HH^2\times\RR^2\times R$, and
not only do these co-ordinates approach limiting values $(z_1,z_2,q_1,q_2,\lambda)$
in $\HB^2\times\RR^2\times\overline R$,
but also the co-ordinates $r_1$, $\omega_1$ and $\omega_2$ defined by \rf{bu1.1} and \rf{bu1.2} and the co-ordinates
$r_2$, $\eta_1$ and $\eta_2$ defined by \rf{bu2.1} and \rf{bu2.2} approach limiting values as well. Of course,
the co-ordinates $r_2$, $\eta_1$ and $\eta_2$ are only defined in the region where $|z_1|$ and $|z_2|$ are bounded.
But, for these co-ordinates, we really care only about the point where $z_i=-\Re(\lambda)+q_i$, $i=1,2$, since
away from the singular locus, the blowup co-ordinates are determined by the base co-ordinates $z_i$, $q_i$ and $\lambda$.

\beginlemmalabel{SigmaDesc}
Let $\Sigma$ be the subset of $K$ where $\mu_2=1$. 
Let $K_0$ denote the subset of $\PA_\infty K$ where 
$\lambda\in(-2\sqrt{2},2\sqrt{2})$,
$q_1=q_2=0$.
%, and $(z_1,z_2)\ne (z_\lambda,z_\lambda)$.    
Then 
\be{conclusion}
\Sigma\cap K_0 = (\Sigma_1\cup \Sigma_2\cup \Sigma_3\cup \Sigma_4)\cap K_0\,,
\ee
where
$$\eqalign{
\Sigma_1 &= \{z_1\ne-\lambda, z_2\ne-\lambda, z_1=z_2\in\PA_\infty\HB, \omega_1=\omega_2\}\,,\cr
\Sigma_2 &= \{z_1 = -\lambda, z_2\ne-\lambda, \omega_1=0\}\,,\cr
\Sigma_3 &= \{z_1\ne-\lambda, z_2 = -\lambda, \omega_2=0\}\,,\cr
\Sigma_4 &= \{z_1 =- \lambda, z_2 = -\lambda, \eta_1=e^{i\psi}\omega_1,\eta_2=e^{i\psi}\omega_2 
{\rm\ for\ some\ } \psi\in[0,\pi]\}\,.\cr
}$$
\endlemma
\beginremark 
In fact we will only use this theorem when both $z_1$ and $z_2$ are in $\PA_\infty\HB$.
\endremark
\beginproof 
Assume for the moment that $(z_1,z_2,q_1,q_2,\lambda)\in \HH^2\times\RR^2\times\overline R$. 
Since
$$
\chi(\phi(z_1,z_2,q_1,q_2,\lambda)) 
= {{\Im(z_1+\lambda)|z_2+\lambda-q_2|^2 + \Im(z_2+\lambda)|z_1+\lambda-q_1|^2}
\over{|z_1 + \lambda - q_1 + z_2 + \lambda - q_2 + z_\lambda(z_1 + \lambda - q_1)(z_2 + \lambda - q_2)|^2}},
$$
the function $\mu_2$ is given by
$$\displaylines{\quad
\mu_2(z_1,z_2,q_1,q_2,\lambda) 
\hfill\cr\hfill
\eqalign{&={{2 \chi(z_1)\chi(z_2)}\over{\chi(\phi(z_1,z_2,\lambda,q_1,q_2))(\chi(z_1)+\chi(z_2))}}\cr
&={{2 \chi(z_1)\chi(z_2)|z_1 + \lambda - q_1 + z_2 + \lambda - q_2 + z_\lambda(z_1 + \lambda - q_1)(z_2 + \lambda - q_2)|^2}\over
{(\Im(z_1+\lambda)|z_2+\lambda-q_2|^2 + \Im(z_2+\lambda)|z_1+\lambda-q_1|^2)(\chi(z_1)+\chi(z_2))}}\cr
&={{2 \chi(z_1)\chi(z_2)|z_1 + \lambda - q_1 + z_2 + \lambda - q_2 + z_\lambda(z_1 + \lambda - q_1)(z_2 + \lambda - q_2)|^2}\over
{(p_1\chi(z_1)|z_1-z_\lambda|^2|z_2 + \lambda - q_2|^2+p_2\chi(z_2)|z_2-z_\lambda|^2|z_1 + \lambda - q_1|^2)(\chi(z_1)+\chi(z_2))}}\,,
}\quad}$$
where
$$
p_i = 1 + \Im(\lambda)/\Im(z_i)\,.
$$
Define $\mu_2^*$ by setting $p_1=p_2=1$ in this formula, that is,
$$\displaylines{\quad
\mu_2^*(z_1,z_2,q_1,q_2,\lambda)
\hfill\cr\hfill\lastdisplayline{mu2star}{
\eqalign{
&=
{{2 \chi(z_1)\chi(z_2)|z_1 + \lambda - q_1 + z_2 + \lambda - q_2 + z_\lambda(z_1 + \lambda - q_1)(z_2 + \lambda - q_2)|^2}\over
{(\chi(z_1)|z_1-z_\lambda|^2|z_2 + \lambda - q_2|^2+\chi(z_2)|z_2-z_\lambda|^2|z_1 + \lambda - q_1|^2)(\chi(z_1)+\chi(z_2))}}\cr
&=
{{2 \omega_1\omega_2|z_1 + \lambda - q_1 + z_2 + \lambda - q_2 + z_\lambda(z_1 + \lambda - q_1)(z_2 + \lambda - q_2)|^2}\over
{(\omega_1|z_1-z_\lambda|^2|z_2 + \lambda - q_2|^2+\omega_2|z_2-z_\lambda|^2|z_1 + \lambda - q_1|^2)(\omega_1+\omega_2)}}\,.\cr
}}\quad}
$$
Clearly $\mu_2 \le \mu_2^*$.

Now let $k\in \Sigma \cap K_0$.
To show the inclusion $\subseteq$ in \rf{conclusion} we must show that $k$ is in $\Sigma_i$ for some $i\in\{1,2,3,4\}$.
Let the co-ordinates of k be given by the base co-ordinates $z_1$, $z_2$, $q_1=0$, $q_2=0$, 
$\lambda\in(-2\sqrt{2}$, $2\sqrt{2})$, the first blow-up co-ordinates $ r_1$, $\omega_1$ and $\omega_2$
and, if $z_1,z_2\ne i\infty$, the second blow-up co-ordinates $r_2$, $\eta_1$ and $\eta_2$. Since $k\in \PA_\infty K$,
$\mu_2(k)$ is defined as a $\limsup$.

The points of continuity of $\mu_2^*$ are the points where the denominator of \rf{mu2star} does not vanish.
Thus $k$ satisfies one of the following four mutually disjoint conditions:\par
(i) $k$ is a point of continuity for $\mu_2^*$,\par
(ii) $z_1=-\lambda$ and $z_2=-\lambda$,\par
(iii) $z_1=-\lambda$, $z_2\ne-\lambda$ and $\omega_1=0$,\par
(iv) $z_1\ne-\lambda$, $z_2=-\lambda$ and $\omega_2=0$.\par
\noindent If conditions (iii) or (iv) hold then $k$ lies in $\Sigma_2$ or $\Sigma_3$ and we are done. 

Suppose that (i) holds. Then
$$
1=\mu_2(k)= \limsup_{{k_n\rarr k}\atop{k_n\in K\backslash \PA_\infty K}}\mu_2(k_n)
\le \limsup_{{k_n\rarr k}\atop{k_n\in K\backslash \PA_\infty K}}\mu_2^*(k_n) \le 1\,.
$$
The last inequality holds because at a point of continuity, the $\limsup$ is actually a limit
which can be evaluated in any order. If we take the limit in $\lambda$ and $q_i$ first, we may 
use the fact that for $\lambda\in(-2\sqrt{2},2\sqrt{2})$, $\mu_2=\mu_2^*$, and from \thmrf{mu2le1} 
we see that the remaining limit in $z_1$ and $z_2$ can be at most $1$. 

Thus we have $\mu_2^*(k)=1$ and we need to show that if (i) holds, and (ii), (iii) and (iv) do not, then
$k$ lies in one of the sets $\Sigma_i$. Let us first consider the case where $z_1=z_2=i\infty$.
In this case we must introduce new variables $w_i=-1/z_i$, substitute into \rf{mu2star} and send
$w_1$ and $w_2$ to zero. Using $|z_\lambda|^2=2$ for $\lambda\in(-2\sqrt{2},2\sqrt{2})$ we find
that at this point $\mu_2^*=4\omega_1\omega_2/(\omega_1+\omega_2)^2$. So $\mu_2^*=1$ implies that
$\omega_1=\omega_2$ and thus that $k\in\Sigma_1$.

Continuing with case (i) let us consider next the possibility that $z_1\in\PA_\infty\HB$ and $z_2\not\in\PA_\infty\HB$.
Then $\omega_1=0$ and $\omega_2=1$ and the numerator in \rf{mu2star} is zero. Away from points described by
(ii), (iii) and (iv) the denominator is not zero so $\mu_2^*(k)=0$. This is impossible. When $z_1=i\infty$ we
should first replace $z_1$ with $-1/w_1$ and send $w_1$ to zero. This leads to the same conclusion.

Thus we are left to consider points satisfying (i) where $z_1$ and $z_2$ are both in $\PA_\infty\HB$ and
not the point at infinity. In other words $z_1$ and $z_2$ are real but not equal to $-\lambda$. In this
case the condition $\mu_2^*=1$ can be rewritten as
\be{Mker}
\twovec{\omega_1}{\omega_2}^T M \twovec{\omega_1}{\omega_2} = 0\,,
\ee
with
$$
M=\twovec{m_{1,1} & m_{1,2}}{m_{2,1} & m_{2,2}}\,,
$$
where $m_{1,2} = m_{2,1}$ and
$$\eqalign{
m_{1,1} &= |z_1-z_\lambda|^2|z_2+\lambda|^2\,,\cr
m_{1,2} &=  (|z_1-z_\lambda|^2|z_2+\lambda|^2+|z_2-z_\lambda|^2|z_1+\lambda|^2)/2 
- |z_1+\lambda + z_2+\lambda+z_\lambda(z_1+\lambda)(z_2+\lambda)|^2\,,\cr
m_{2,2} &= |z_2-z_\lambda|^2|z_1+\lambda|^2\,.\cr
}$$
Setting $s_1=z_1+\lambda$ and $s_2=z_2+\lambda$ and using $s_1, s_2 \in \RR$ we find
$$
M = \twovec
{(s_1^2-\lambda s_1 + 2)s_2^2 & -s_1s_2(s_1s_2-\lambda(s_1+s_2)/2 + 2}
{ -s_1s_2(s_1s_2-\lambda(s_1+s_2)/2 + 2 & (s_2^2-\lambda s_2 + 2)s_1^2}.
$$
Since $\tr(M)\ge 0$ the condition \rf{Mker} requires
$$
\det(M) = s_1^2s_2^2(s_1-s_2)^2(2-\lambda^2/4) = 0\,.
$$
But $s_1$ and $s_2$ are not zero. Therefore we conclude that $s_1=s_2$ and thus $z_1=z_2$. In addition,
$$
M=s^2(s^2-\lambda s + 2)\twovec{1&-1}{-1&1}\,,
$$
where $s$ is the common value of $s_1$ and $s_2$. For $\lambda\in(-2\sqrt{2},2\sqrt{2})$, $s^2-\lambda s + 2>0$.
Thus \rf{Mker} implies that $\omega_1=\omega_2$
and so $k\in\Sigma_1$.

Finally, we must deal with case (ii). 
Let $(z_{1,n},z_{2,n},q_{1,n},q_{2,n},\lambda_n) \in \HH^2\times\RR^2\times\overline R$ be a sequence
that realizes the $\limsup$ in the definition of $\mu_2(k)$. Define $\tilde r_{2,n}$,
$\tilde\eta_{1,n}$ and $\tilde\eta_{2,n}$ via
$$\eqalign{
z_{1,n} + \lambda_n - q_{1,n} &= \tilde r_{2,n}\tilde\eta_{1,n}\,,\cr
z_{2,n} + \lambda_n - q_{2,n} &= \tilde r_{2,n}\tilde\eta_{2,n}\,,\cr
}$$
and
$$
|\tilde \eta_{1,n}|^2 + |\tilde \eta_{2,n}|^2 =1\,.
$$
By going to a subsequence if needed, we may assume that $\tilde r_{2,n}$, $\tilde\eta_{1,n}$ and $\tilde\eta_{2,n}$ converge
to $0$, $\tilde\eta_{1}$ and $\tilde\eta_{2}$ respectively. 
We may also assume that $p_{i,n} = 1 + \Im(\lambda_n)/\Im(z_{i,n})$ converge
to $p_i$ for $i=1,2$. 

Then we find that 
\be{premove}
1=\mu_2(k) 
= {{\omega_1\omega_2|\tilde\eta_1+\tilde\eta_2|^2}
\over{(p_1\omega_1|\tilde\eta_2|^2+p_2\omega_2|\tilde\eta_1|^2)(\omega_1+\omega_2)}}
\le{{\omega_1\omega_2|\tilde\eta_1+\tilde\eta_2|^2}
\over{(\omega_1|\tilde\eta_2|^2+\omega_2|\tilde\eta_1|^2)(\omega_1+\omega_2)}}
\le 1\,,
\ee
unless the denominator is zero, that is, $\omega_1=0$, $\tilde\eta_1=0$ or 
$\omega_2=0$, $\tilde\eta_2=0$, which we assume for the moment is not the case.
The last inequality holds because it is equivalent to
$$
\twovec{\omega_1}{\omega_2}^T
\twovec{|\tilde\eta_2|^2 & -\Re(\overline{\tilde\eta_1}\tilde\eta_2)}
{-\Re(\overline{\tilde\eta_1}\tilde\eta_2) & |\tilde\eta_1|^2}
\twovec{\omega_1}{\omega_2} \ge 0\,,
$$
and the matrix in this formula is positive semi-definite. 

Still under the assumption that neither $\omega_1=0$, $\tilde\eta_1=0$ nor 
$\omega_2=0$, $\tilde\eta_2=0$ hold, we see that at least one of $p_1$ or $p_2$
must equal $1$. Otherwise we would have a strict inequality in \rf{premove} which
is impossible. If $p_i=1$ then $\Im(\lambda_n)/\Im(z_{i,n}) \rarr 0$. This
implies that $\tilde r_{2,n}/ r_{2,n} \rarr 1$, because $r_{2,n} \ge \Im(z_{i,n})$
implies $\Im(\lambda_n)/r_{2,n} \rarr 0$ and
$$
\tilde r_{2,n}^2 = \sum_{i=1}^{2}\Re(z_{i,n}-q_{i,n}+\lambda_n)^2 + \Im(z_{i,n}+\lambda_n)^2\,.
$$
Now, from $\tilde r_{2,n}\tilde\eta_{i,n} = r_{2,n} \eta_{i,n} + \Im(\lambda_n)$ we conclude that
$\tilde\eta_{i,n}$ and $\eta_{i,n}$ have the same limit $\eta_i$. So, in fact, we have that
\rf{Mker} holds with
$$
M = \twovec{|\eta_2|^2 & -\Re(\overline\eta_1\eta_2)}
{-\Re(\overline\eta_1\eta_2) & |\eta_1|^2}\,.
$$
Since $\tr(M)> 0$ this requires 
$$
\det(M) = |\eta_1|^2|\eta_2|^2\big(1-\cos(2(\arg(\eta_1)-\arg(\eta_2)))\big) = 0\,.
$$
This means either $\eta_1=0$, $\eta_2=0$, $\arg(\eta_1)=\arg(\eta_2)$ or
$\arg(\eta_1)=\arg(\eta_2) +\pi$. If $\eta_1=0$ then $\twovec{\omega_1}{\omega_2}\in\Ker(M)$
requires $\omega_1=0$ and $k\in\Sigma_{4}$. Similarly, if $\eta_2=0$
then $k\in\Sigma_{4}$. If $\arg(\eta_1)=\arg(\eta_2)=\psi$ then
$$
M=\twovec{|\eta_2|^2 & -|\eta_1||\eta_2|}{-|\eta_1||\eta_2| & |\eta_1|^2}
$$
and \rf{Mker} and the fact that $\omega_1,\omega_2 \ge 0$
implies that $\twovec{\omega_1}{\omega_2} = \twovec{|\eta_1|}{|\eta_2|}$.
Thus $\eta_1=e^{i\psi}\omega_1$ and $\eta_2=e^{i\psi}\omega_2$, and
again we have $k\in\Sigma_4$.
The remaining possibility is that $\arg(\eta_1)=\arg(\eta_2) +\pi$. Since
both $\eta_1$ and $\eta_2$ lie in the upper half plane, this implies that
they are both real with opposite signs. Equation \rf{Mker} then requires $\omega_1=\eta_1$
and $\omega_2=\eta_2$. But this is impossible, as $\omega_1$ and $\omega_2$ are both
non-negative.

To complete the proof we must return to the possibility that 
$\omega_1=0$, $\tilde\eta_1=0$ or $\omega_2=0$, $\tilde\eta_2=0$.
Clearly, at most one of these can hold. Suppose $\omega_1=0$, $\tilde\eta_1=0$.
(The other possibility is handled similarly.)
Introduce one more set of variables $s_n$, $\alpha_{1,n}$ and $\alpha_{2,n}$
satisfying
$$\eqalign{
\omega_{1,n} &= s_{n}^2\alpha_{1,n}\,,\cr
\tilde\eta_{1,n} &= s_{n}^2\alpha_{2,n}\cr
}$$
and
$$
\alpha_{1,n}^2 + |\alpha_{2,n}|^2 = 1\,.
$$
Then $s_n\rarr 0$ and going to a subsequence we may assume that $\alpha_{1,n}$
and $\alpha_{2,n}$ converge to $\alpha_1$ and $\alpha_2$. Then 
$\mu_2(k) = \alpha_1/(p_1\alpha_1+p_2|\alpha_2|^2) = 1$ so that $\alpha_2=0$ 
and $p_1=1$. But $p_1=1$ implies $\tilde\eta_1 = \eta_1$ by the argument above.
Thus $\eta_1=\tilde\eta_1=0$ and $\omega_1=0$ which implies that $k\in\Sigma_4$.
\endproof

%\vfill\eject
\hd{Proofs of \thmrf{mu3bdd} and \thmrf{mu3qbdd}}

\beginproofof{\thmrf{mu3bdd}}
Extend $\mu_{3,p}$ to an upper semi-continuous function 
on $\HB\,^3\times\RR^4\times\RB$ by setting, at points
$Z_0,Q_0,\lambda_0$ where it is not already defined,
$$
\mu_{3,p}(Z_0,Q_0,\lambda_0) = \limsup_{Z\rarr Z_0,Q\rarr Q_0,\lambda\rarr\lambda_0}\mu_{3,p}(Z,Q,\lambda)\,.
$$
Here the limsup is taken over points in $\HH^3\times\RR^4\times R$,
and we are using the notation $Z=(z_1,z_2,z_3)$ for points in $\HH^3$
and $Q=(q_1,q_2,q_3,q_4)$ for points in $\RR^4$. The points $Z$, $Q$ and $\lambda$ are approaching
their limits in the topology of $\HB^3\times\RR^4\times\RB$.

To prove the theorem it is then enough to show that 
\be{cptbdd}
\mu_{3,p}(Z,Q,\lambda) < 1
\ee
for $(Z,Q,\lambda)$ in the compact set $\PA_\infty(\HB\,^3)\times\{0\}^4\times [-E,E]$,
since this implies that for some $\epsilon>0$, the upper semi-continuous function
$\mu_{3,p}(Z,Q,\lambda)$ is bounded by
$1-2\epsilon$ on
the set, and by $1-\epsilon$ in some neighbourhood. 

We will rewrite $\mu_{3,p}$ in terms of 
the simpler function $\mu_2$. 
Define
$$
\nu_i(Z) = {{\cd(z_i)}\over{\cd(z_1)+\cd(z_2)+\cd(z_3)}}\,,
$$
and the maps $\xi_\sigma$ and $\tau_\sigma$ from 
$\HH^3 \times \RR^4\times R$ to $\HH\times\HH\times \RR^2\times L$
labelled by a permutation $\sigma$ of $(1,2,3)$ and given by
$$\eqalign{
\xi_{\sigma}(Z,Q,\lambda)&= (z_{\sigma_2},z_{\sigma_3},q_{\sigma_2},q_{\sigma_3} ,\lambda)\,,\cr
\tau_{\sigma}(Z,Q,\lambda)&= (z_{\sigma_1},\phi(z_{\sigma_2},z_{\sigma_3},q_{\sigma_2},q_{\sigma_3},\lambda),q_{\sigma_1},q_4,\lambda)\,.\cr
}$$
Then we have
$$\displaylines{\quad
\mu_{3,p}(Z,Q,\lambda) 
\hfill\cr\hfill
\lastdisplayline{32formula}{
\eqalign{
&=\sum_{\sigma} {{\cd^p(\phi(z_{\sigma_1}, \phi(z_{\sigma_2}, z_{\sigma_3},
q_{\sigma_2},q_{\sigma_3}, \lambda), q_{\sigma_1},q_{\sigma_4},\lambda))}\over{
\cd^p(z_1) + \cd^p(z_2) + \cd^p(z_3) }}\cr
&=\sum_{\sigma}
\left(
{{\cd(\phi(z_{\sigma_1}, \phi(z_{\sigma_2}, z_{\sigma_3},
q_{\sigma_2},q_{\sigma_3}, \lambda), q_{\sigma_1},q_{\sigma_4},\lambda))}
\over{\cd(z_1) + \cd(z_2) + \cd(z_3)}}
\right)^p
{{1}\over{\nu_1^p+\nu_2^p+\nu_3^p}}\cr
&=\sum_{\sigma}
\left(
\mu_2(\tau_\sigma(Z,Q,\lambda))
\left({{1}\over{2}}\nu_{\sigma_1} 
+ {{1}\over{4}}\mu_2(\xi_\sigma(Z,Q,\lambda))(\nu_{\sigma_2}+\nu_{\sigma_3})\right)
\right)^p
{{1}\over{\nu_1^p+\nu_2^p+\nu_3^p}}\,.\cr
}}\quad}$$

Let $R_1$, $\Omega_1$, $\Omega_2$ and $\Omega_3$ be three dimensional
polar co-ordinates defined as functions of $Z=(z_1,z_2,z_3)\in\HH^3$  by
$$\eqalign{
\chi(z_1) &= R_1 \Omega_1\,, \cr
\chi(z_2) &= R_1 \Omega_2\,, \cr
\chi(z_3) &= R_1 \Omega_3 \cr
}$$
and
$$
\Omega_1^2 + \Omega_2 ^2 + \Omega_3 ^2 =1\,.
$$
Notice that for any permutation $\sigma$ of $(1,2,3)$,
\be{nuOmega}
\nu_{\sigma_1}  = {{\Omega_{\sigma_2} \Omega_{\sigma_3} }
\over{\Omega_1 \Omega_2 +\Omega_1 \Omega_3 +\Omega_2 \Omega_3 }}\,.
\ee

Next, let $r_1(z_1,z_2)$, $\omega_1(z_1,z_2)$ and $\omega_2(z_1,z_2)$ be the co-ordinates
defined by \rf{bu1.1} and $\rf{bu1.2}$. Then, for any permutation $\sigma$ of $(1,2,3)$ and
any $Z=(z_1,z_2,z_3)\in\HH^3$,
\be{Oo}\eqalign{
\Omega_{\sigma_2} ^2 &= \left(\Omega_{\sigma_2} ^2+\Omega_{\sigma_3} ^2\right)
\omega_1^2(z_{\sigma_2},z_{\sigma_3})\,,\cr
\Omega_{\sigma_3} ^2 &= \left(\Omega_{\sigma_2} ^2+\Omega_{\sigma_3} ^2\right)
\omega_2^2(z_{\sigma_2},z_{\sigma_3})\,,\cr
}\ee
where each $\Omega_{\sigma_i}$ is evaluated at $Z$. To see this note that, since $\chi(z_{\sigma_2}) 
= r_1(z_{\sigma_2},z_{\sigma_3})\omega_1(z_{\sigma_2},z_{\sigma_3})
= R_1 \Omega_{\sigma_2} $ and $\chi(z_{\sigma_3})=r_1(z_{\sigma_2},z_{\sigma_3})\omega_2(z_{\sigma_2},z_{\sigma_3})
= R_1 \Omega_{\sigma_3} $ we have 
$$
r_1^2(z_{\sigma_2},z_{\sigma_3})
= r_1^2(z_{\sigma_2},z_{\sigma_3})
(\omega_1^2(z_{\sigma_2},z_{\sigma_3}) + \omega_2^2(z_{\sigma_2},z_{\sigma_3}))
= R_1 ^2(\Omega_{\sigma_2} ^2 + \Omega_{\sigma_3} ^2) \,.
$$
Thus
$$
R_1 ^2\Omega_{\sigma_2} ^2
=r_1^2(z_{\sigma_2},z_{\sigma_3})\omega_1^2(z_{\sigma_2},z_{\sigma_3})
=R_1 ^2(\Omega_{\sigma_2} ^2 + \Omega_{\sigma_3} ^2)\omega_1^2(z_{\sigma_2},z_{\sigma_3})
$$
and since $R_1  \ne 0$ for $Z\in\HH^3$ the first equality of \rf{Oo} follows.
The second equality is proved in the same way. A similar argument also shows
\be{OF}\eqalign{
\Omega_{\sigma_1} ^2 &=
\left(\Omega_{\sigma_1} ^2 + 
F^2
(\Omega_{\sigma_2} ^2+\Omega_{\sigma_3} ^2)\right)
\omega_1^2(z_{\sigma_1},\phi(z_{\sigma_2},z_{\sigma_3},q_{\sigma_2},q_{\sigma_3},\lambda)) \,,
\cr
F^2
(\Omega_{\sigma_2} ^2+\Omega_{\sigma_3} ^2)
&=
\left(\Omega_{\sigma_1} ^2 + 
F^2
(\Omega_{\sigma_2} ^2+\Omega_{\sigma_3} ^2)\right)
\omega_2^2(z_{\sigma_1},\phi(z_{\sigma_2},z_{\sigma_3},q_{\sigma_2},q_{\sigma_3},\lambda))\,.\cr
}\ee
Here each $\Omega_{\sigma_i}$ is evaluated at $Z$ and 
\be{Fformula}\eqalign{
F &= F(z_{\sigma_2},z_{\sigma_3},q_{\sigma_2},q_{\sigma_3},\lambda)\cr 
&= 
{{\chi(\phi(z_{\sigma_2},z_{\sigma_3},q_{\sigma_2},q_{\sigma_3},\lambda))}\over{r_1(z_{\sigma_2},z_{\sigma_3})}}\cr
&={{2\omega_1(z_{\sigma_2},z_{\sigma_3})\omega_2(z_{\sigma_2},z_{\sigma_3})}\over{\mu_2(z_{\sigma_2},z_{\sigma_3},q_{\sigma_2},q_{\sigma_3},\lambda)(\omega_1(z_{\sigma_2},z_{\sigma_3})+\omega_2(z_{\sigma_2},z_{\sigma_3}))}}\,.\cr
}\ee
We will prove \rf{cptbdd} by contradiction. For this suppose that $\mu_{3,p}(Z,Q,\lambda) = 1$
for some $(Z,Q,\lambda)\in\PA_\infty(\HB\,^3)\times\{0\}^4\times [-E,E]$. Then there must
exist a sequence $(Z_n,Q_n,\lambda_n)$ with $Z_n\rarr Z$ in $\HB\,^3$ $Q_n\rarr(0,0,0,0)$ and
$\lambda_n\rarr\lambda\in[-E,E]$ such that
$$
\lim \mu_{3,p}(Z_n,Q_n,\lambda_n) = 1\,.
$$
From now on $Z=(z_1,z_2,z_3)$ and $\lambda$ will denote the 
limiting values of the sequence $Z_n$ and $\lambda_n$.
Similarly, we will denote by $\nu_i$ and $\Omega_i$ the limits of $\nu_i(Z_n)$
and $\Omega_i(Z_n)$. 
We claim that
\be{nuequal}
\nu_1=\nu_2=\nu_3 = {{1}\over{3}}\,.
\ee
This follows from \rf{32formula}, the bound $\mu_2\le 1$ proved in
\thmrf{mu2le1}
and convexity of $x\mapsto x^p$ which imply
$$\eqalign{
1 
&\le \sum_{\sigma}
\left({{1}\over{2}}\nu_{\sigma_1} 
+ {{1}\over{4}}(\nu_{\sigma_2}+\nu_{\sigma_3})\right)^p
{{1}\over{\nu_1^p+\nu_2^p+\nu_3^p}}\cr
&\le  \sum_{\sigma}
\left({{1}\over{2}}\nu_{\sigma_1}^p
+ {{1}\over{4}}(\nu_{\sigma_2}^p+\nu_{\sigma_3}^p)\right)
{{1}\over{\nu_1^p+\nu_2^p+\nu_3^p}}\cr
&=1\,,
}$$
so the inequalities must actually be equalities. Since $p>1$, 
strict convexity implies that equality only holds if $\nu_1=\nu_2=\nu_3$. 
Since their sum is $1$, their common value must be $1/3$.

By going to a subsequence, we may assume that $\Omega_i(Z_n)$ converge. 
Then \rf{nuequal} and \rf{nuOmega} imply that their limiting values along
the sequence must be 
\be{OmegaEqual}
\Omega_1=\Omega_2=\Omega_3=1/\sqrt{3}\,.
\ee
One consequence is that
\be{zinHB}
z_i \in \PA_{\infty} \HB
\ee
for $i=1,2,3$.

Now consider the values of $\xi_\sigma(Z_n,Q_n,\lambda_n)$ and 
$\tau_\sigma(Z_n,Q_n,\lambda_n)$. Since these vary in a compact region in $K$
we may, again by going to a subsequence, assume that they converge in $K$ 
to values which we will denote
$\xi_\sigma$ and $\tau_\sigma$. Returning to \rf{32formula} and using
\rf{nuequal}, the upper semi-continuity of $\mu_2$ and the bound $\mu_2\le 1$, 
we find that
$$
\eqalign{
1 &= \lim_{n\rarr\infty} {{1}\over{3}}
\sum_\sigma \left({{\mu_2(\tau_\sigma(Z_n,Q_n,\lambda_n))(1+\mu_2(\xi_\sigma(Z_n,Q_n,\lambda_n)))}\over{2}}\right)^p\cr
&\le {{1}\over{3}}\sum_\sigma\left({{\mu_2(\tau_\sigma)(1+\mu_2(\xi_\sigma))}\over{2}}\right)^p\cr
&\le 1\,.\cr
}$$
This implies that for every $\sigma$ occurring in the sum we have
$$
\mu_2(\xi_\sigma) = \mu_2(\tau_\sigma) = 1\,.
$$
This and \rf{zinHB} imply that for each $\sigma$, $\xi_\sigma$ and $\tau_\sigma$ lie in the set $\Sigma$ of 
\thmrf{SigmaDesc}.

Now consider the co-ordinates $\omega_1$ and $\omega_2$ for the the point $\xi_\sigma$. 
These are the limiting values of $\omega_i(z_{\sigma_2},z_{\sigma_3})$ along our sequence.
Equations \rf{Oo} and \rf{OmegaEqual} then imply that these limiting values are 
$\omega_1=\omega_2=1/\sqrt{2}$. Examining the description of $\Sigma$ in \thmrf{SigmaDesc}, we
conclude that the $\HB$ co-ordinates of $\xi_\sigma$, namely the limiting values of
$z_{\sigma_2}$ and $z_{\sigma_3}$, must be equal. Since this is
true for every $\sigma$ we conclude that
$$
z_1=z_2=z_3 \in \PA_\infty\HB\,.
$$
Let $z$ denote their common value.

We first consider the possibility that $z \ne -\lambda$. The two $\HB$ co-ordinates of
the point $\tau_{(1,2,3)}$ are $z$ and the limiting value of 
$\phi(z_2,z_3,q_2,q_3,\lambda)$. This limiting value
is simply $\phi(z,z,0,0,\lambda) = -2/(z+\lambda)$ and is easily seen to be not equal to $z$.
The only way that $\tau_\sigma$ with $\HB$ co-ordinates $z$ and $\phi(z,z,0,0,\lambda)$ can
lie in $\Sigma$ with $z\ne-\lambda$ is that $\phi(z,z,0,0,\lambda)=-\lambda$ and that
the $\omega_2$ co-ordinate is $0$. The $\omega_2$ co-ordinate is the limiting value
of $\omega_2(z_1,\phi(z_2,z_3,q_2,q_3,\lambda))$ which we may use in taking the limit
of equation \rf{OF}. The limiting value of $F$ in that equation can be computed from
\rf{Fformula}, since we know that the values of $\omega_i$ in that formula are $1/\sqrt{2}$
and the value of $\mu_2$ in that formula is $1$. This gives $F=1/\sqrt{2}$ and so the 
second equation of \rf{OF} yields
$1/3 = 0$ in the limit, which is impossible.

This leaves the possibility that $z = -\lambda$. Again, the $\HB$ co-ordinates for the point
$\tau_{(1,2,3)}$ are $z$ and the limiting value of 
$\phi(z_2,z_3,q_2,q_3,\lambda)$. By going to a subsequence, we have
assumed that this limiting value exists.
However, in this case it is not clear what the value is, since $(-\lambda,-\lambda,0,0,\lambda)$
is the point where $\phi$ is not continuous. In fact, we will see that the limiting
value, possibly after going one more time to a subsequence, is $i\infty$. To see this
we write
$$\eqalign{
\phi(z_{2,n},z_{3,n},q_{2,n},q_{3,n},\lambda_n)
&={{-(z_{2,n}+\lambda_n-q_{2,n})-(z_{3,n}+\lambda_n-q_{3,n})}\over
{(z_{2,n}+\lambda_n-q_{2,n})(z_{3,n}+\lambda_n-q_{3,n})}}\cr
&={{-r_{2,n}(\eta_{1,n}+\eta_{2,n}) - 2i\Im(\lambda_n)}
\over{(r_{2,n}\eta_{1,n}+i\Im(\lambda_n))(r_{2,n}\eta_{2,n}+i\Im(\lambda_n))}}\,,
}$$
where $r_{2,n}$, $\eta_{1,n}$ and $\eta_{1,n}$ are the co-ordinates defined by
\rf{bu2.1} and \rf{bu2.2}. Since for our sequence, $z_{2,n},z_{3,n}\rarr-\lambda$, 
$q_{2,n},q_{3,n}\rarr 0$ we have $r_{2,n}\rarr 0$. We also have $\Im\lambda_n\rarr 0$
so if we write $(r_{2,n}, \Im(\lambda_n))$ in polar co-ordinates, that is,
$r_{2,n} = s_n\alpha_{1,n}$ and $\Im(\lambda_n)=s_n\alpha_{2,n}$ with $\alpha_{1,n}^2+\alpha_{2,n}^2=1$,
then $s_n\rarr 0$. By going to a subsequence we may assume $\alpha_{1,n}$ and $\alpha_{2,n}$ converge
to non-negative values $\alpha_1$ and $\alpha_2$. Then
$$
\phi(z_{2,n},z_{3,n},q_{2,n},q_{3,n},\lambda_n) = 
{{-\alpha_{1,n}(\eta_{1,n}+\eta_{2,n}) - 2i\alpha_{2,n}}
\over{s_n(\eta_{1,n}+i\alpha_{1,n})(\eta_{2,n}+i\alpha_{2,n})}}\,.
$$
The denominator of this expression converges to $0$. The numerator converges to
$-\alpha_1(\eta_1+\eta_2) - 2i\alpha_2$ where $\eta_1$ and $\eta_2$ are co-ordinates
in $K$ for $\xi_{(1,2,3)}$. Since $\xi_{(1,2,3)}$ lies in $\Sigma$ with $\HB$ co-ordinates
$(-\lambda, -\lambda)$ we must have $\eta_1+\eta_2=e^{-i\psi}\sqrt{2}$ with
$\psi\in[0,\pi]$. Here we used that the $\omega$ co-ordinates for $\xi_{(1,2,3)}$ are both
$1/\sqrt{2}$. But now we see that it is impossible that $\alpha_1(\eta_1+\eta_2) + 2i\alpha_2=0$,
since that imaginary part being zero forces $\alpha_2=0$ and $\psi\in\{0,\pi\}$ in which case
$\alpha_1=1$ so that $\alpha_1(\eta_1+\eta_2) + 2i\alpha_2=\pm\sqrt{2}\ne 0$. This implies
that the limiting value of $\phi$ is $i\infty$.

Now we know that the point $\tau_{(1,2,3)}$ has $\HB$ co-ordinates $-\lambda$ and $i\infty$. Thus
$\tau_{(1,2,3)}\in\Sigma$ requires that the $\omega_1$ co-ordinate of $\tau_{(1,2,3)}$
be zero. Arguing as above, we find that in the limit, the first 
equation of \rf{OF} reads $1/3 = 0$. This contradiction concludes the proof of the theorem.
\endproof

\bigskip

\beginproofof{\thmrf{mu3qbdd}}
Each term in the sum appearing in $\mu_{3,p}$ can be estimated
$$\eqalign{
{{\cd^p(\phi(\cdots\cdots))}\over{\cd^p(z_1) + \cd^p(z_2) + \cd^p(z_3)}}
&={{(\cd(z_1) + \cd(z_2) + \cd(z_3))^p}\over{\cd^p(z_1) + \cd^p(z_2) + \cd^p(z_3)}}
\left({{\cd(\phi(\cdots\cdots))}\over{\cd(z_1) + \cd(z_2) + \cd(z_3)}}\right)^p\cr
&\le 3^{p-1}\left({{\cd(\phi(\cdots\cdots))}\over{\cd(z_1) + \cd(z_2) + \cd(z_3)}}\right)^p,\cr
}$$
where $\phi(\cdots\cdots)$ denotes $\phi(z_{\sigma_1}, \phi(z_{\sigma_2}, z_{\sigma_3},
q_{\sigma_2},q_{\sigma_3}, \lambda), q_{\sigma_1},q_4,\lambda)$.
Therefore it is enough to prove
\be{pone}
{{\cd(\phi(\cdots\cdots))}\over{\cd(z_1) + \cd(z_2) + \cd(z_3)}} \le C(1+\sum_{i=1}^4 |q_i|^2)\,.
\ee
Let $\phi(\cdots)$ denote $\phi(z_{\sigma_2}, z_{\sigma_3},
q_{\sigma_2},q_{\sigma_3}, \lambda)$. Then
$$\eqalign{
\Im(\phi(\cdots)) 
&= 
{{\Im(z_{\sigma_2}+\lambda)}\over{|z_{\sigma_2}+\lambda-q_{\sigma_2}|^2}}
+
{{\Im(z_{\sigma_3}+\lambda)}\over{|z_{\sigma_3}+\lambda-q_{\sigma_3}|^2}}\cr
&\ge 
{{\Im(z_{\sigma_2})}\over{|z_{\sigma_2}+\lambda-q_{\sigma_2}|^2}}\,.\cr
}$$
Thus we have
$$\displaylines{\quad
{{\cd(\phi(\cdots\cdots))}\over{\cd(z_1) + \cd(z_2) + \cd(z_3)}}
\hfill\cr\hfill\eqalign{
&=
\displaystyle{{
\left|
-\displaystyle{{1}\over{z_{\sigma_1}+\lambda-q_{\sigma_1}}}
-\displaystyle{{1}\over{\phi(\cdots)+\lambda-q_{\sigma_4}}}-z_\lambda
\right|^2
}\over{
\Im\left(
-\displaystyle{{1}\over{z_{\sigma_1}+\lambda-q_{\sigma_1}}}
-\displaystyle{{1}\over{\phi(\cdots)+\lambda-q_{\sigma_4}}}
\right)
}}
\displaystyle{{
1
}\over{
\sum_{i=1}^3 |z_i-z_\lambda|^2 / \Im(z_i)
}}\cr
&=
{{
\left|
(z_{\sigma_1}+\lambda-q_{\sigma_1}) 
+(\phi(\cdots)+\lambda-q_{\sigma_4})
+z_\lambda (z_{\sigma_1}+\lambda-q_{\sigma_1}) (\phi(\cdots)+\lambda-q_{\sigma_4})
\right|^2
}\over{
\Im(z_{\sigma_1}+\lambda)|\phi(\cdots)+\lambda-q_{\sigma_4}|^2
+\Im(\phi(\cdots)+\lambda)|z_{\sigma_1}+\lambda-q_{\sigma_1}|^2
}}\cr
&\quad\quad\times
\displaystyle{{
1
}\over{
\sum_{i=1}^3 |z_i-z_\lambda|^2 / \Im(z_i)
}}\cr
&\le
\left(
{{3
}\over{
\Im(\phi(\cdots))
}}
+
{{ 3 + 3|z_{\sigma_1}+\lambda-q_{\sigma_1}|^2
}\over{
\Im(z_{\sigma_1})
}}
\right)
\displaystyle{{
1
}\over{
\sum_{i=1}^3 |z_i-z_\lambda|^2 / \Im(z_i)
}}\cr
&\le
\left(
{{3|z_{\sigma_2}+\lambda-q_{\sigma_2}|^2
}\over{
\Im(z_{\sigma_2})
}}
+
{{ 3 + 3|z_{\sigma_1}+\lambda-q_{\sigma_1}|^2
}\over{
\Im(z_{\sigma_1})
}}
\right)
\displaystyle{{
1
}\over{
\sum_{i=1}^3 |z_i-z_\lambda|^2 / \Im(z_i)
}}\,.\cr
}\quad}$$
Choose the compact set $K$ so that $\sum_{i=1}^3 |z_i-z_\lambda|^2 / \Im(z_i)\ge C >0$ for
some constant $C$ if $(z_1,z_2,z_3)\in K^c$. 
Then we can estimate each term depending on whether $z_{\sigma_i}$ is close to
$z_\lambda$. If it is sufficiently close, then $\Im(z_{\sigma_i})$ is bounded below
and $|z_{\sigma_i}|$ is bounded above by a constant. Thus
$$
\Im(z_{\sigma_i})\sum_{i=1}^3 |z_i-z_\lambda|^2 / \Im(z_i)\ge \Im(z_{\sigma_i})C \ge C' > 0
$$
and $|z_{\sigma_i}+\lambda-q_{\sigma_i}|^2 \le C(1 + |q_{\sigma_i}|^2)$, so we are done. 
Otherwise
$$
\Im(z_{\sigma_i})\sum_{i=1}^3 |z_i-z_\lambda|^2 / \Im(z_i) \ge |z_{\sigma_i}-z_\lambda|^2
\ge C(1+|z_{\sigma_i}|^2)
$$
so that $|z_{\sigma_i}+\lambda-q_{\sigma_i}|^2 / \left(\Im(z_{\sigma_i})\sum_{i=1}^3 |z_i-z_\lambda|^2/ \Im(z_i)\right)
\le C(1 + |q_{\sigma_i}|^2)$ in this case too.

The estimates for $\mu'_{3,p}$ and $\mu'_{1,p}$ are very similar. We omit the details.
\endproof